\documentclass[11pt]{article}
\usepackage{epsfig}
\usepackage{oldgerm}
\usepackage{cite}
\usepackage{amssymb}
\usepackage{subfigure}
\usepackage{picinpar}
\usepackage{float}
\usepackage{chapterbib}
\usepackage{bibspacing}

\normalsize
\newtheorem{theorem}{Theorem}

\newtheorem{lemma}{Lemma}
\def \MM {\hbox{$\cal M$}}

\def\pr{\vskip \parskip}
\font\tenbit=cmbxti10
\font\sevenbit=cmbxti7
\font\fivebit=cmbxti7 at 5pt
\newfam\bitfam
\textfont\bitfam=\tenbit
\scriptfont\bitfam=\sevenbit
\scriptscriptfont\bitfam=\fivebit
\def\mbit{\fam\bitfam\tenbit}

\let \dpar=\partial
\def \ve#1{{\mbit #1}}
\def \id {\equiv}
\def \df {\hbox{\vbox{\offinterlineskip \hbox to 6mm{\ce{\miV def}}
         \kern .6mm \hbox to 6mm{\ce{=}} }} } %symbol of "def"
\font\sfX  =cmss10 at 10pt % sans-serif
\font\Rmath=msbm10

\font\bfX=cmbx10

\font\bslX=cmbxsl10 at 10pt

\font\bslX=cmbxsl10 at 10pt
\font\slX=cmsl10 at 10pt

\font\bsl=cmbxsl10
\font\miV =cmmi5
\font\rmX=cmr10 at 10pt %smaller roman
\font\rmIX=cmr10 at 9pt %smaller roman
\font\itX=cmti10 at 10pt %smaller italics
\font\itIX=cmti10 at 9pt %smaller italics
\font\slIX=cmsl10 at 9pt %smaller slanted
\def \diff {di\kern -1pt f\kern -1.5pt f}

\def \ll {\hbox{$\lambda$}}
\def \mm {\hbox{$\mu$}}
\def \pp {\hbox{$\pi$}}
\def \Ph {\hbox{$\Phi$}}

\def \th {\hbox{$\theta$}}

\font\got=eufm10
\font\Got=eufm10 at 12pt
\def\BB {\hbox{$\cal B$}}
\def\CC {\hbox{$\cal C$}}
\def\SS {\hbox{$\cal S$}}

\def\GG {\hbox{$\cal G$}}
\def\KK {\hbox{$\cal K$}}
\def\MM {\hbox{$\cal M$}}
\def\kk {\hbox{\got k}}
\def\kL {\hbox{\Got k}}

\def \qqr {\hskip .5em }
\def\blacksq#1%          produces a  black square: \blacksq{2}
{\vrule height #1cm depth 0cm width #1cm}
\def\bibb#1#2{\dimen2=\hsize \advance \dimen2 by -1cm%
\hbox to \hsize{\hbox{\bf [#1]}\hfill\vtop{\hsize=\dimen2%
\parindent 0cm {#2\pr\null}}}}%
\def\ce#1{\hfil#1\hfil}
\def\cel#1{\hbox to \hsize{\hfil#1\hfil}}

\def \Real {\hbox{\Rmath R}}
\def \Natural{\hbox{\Rmath N}}
\def \Om {\Omega}
\def \dOm {\dpar\Omega}
\def \vx { \ve{x}}
\def \wt {\widetilde}

\begin{document}

\null \vskip 0.8 truecm

{\bf
\centerline{\ce {An integral equation method for the inverse conductivity
problem.}}}

\medskip
{\sl \centerline{\ce {S. Ciulli$^{\bf \dagger}$,
M.K. Pidcock$^{\bf \ddagger}$
and C. Sebu$^{\bf \dagger}$}}}

\vskip 0.1cm
{\leftskip=1cm
\noindent $^{\bf \dagger}${\slX Laboratoire de Physique--Math\'ematique
et Th\'eorique, Universit\'e de Montpellier II,}

\noindent \hskip 0.2cm {\slX 34095 Montpellier, France;} \par}

{\leftskip=1cm
\noindent $^{\bf \ddagger}${\slX Department of
Mathematical Sciences,
Oxford Brookes University, Oxford OX33 1HX, }

\noindent \hskip 0.2cm {\slX  United Kingdom.}
%%\hfill [Preliminary version, \Date] }
%\hfill [Preliminary version, July 07, 2003] }
\par}

{\hfill { PM 04-04}}

{\hfill {\it To be published in Physics Letters A}}

\medskip

\noindent {\bslX Abstract:}
{\slIX We present an image reconstruction algorithm for the Inverse
Conductivity Problem based on reformulating the problem in terms of
integral equations. We  use as  data the values of injected electric currents and of the corresponding
induced boundary potentials, as well as the boundary values of the electrical
conductivity.

\noindent
We have used {\itIX a priori} information to  find a
{\itIX regularized}  conductivity distribution by first solving a Fredholm
integral equation of the second kind for the Laplacian of the potential,
and then by solving a first order partial differential equation for the
regularized  conductivity  itself. Many of the
calculations involved in the method can be achieved analytically using
the eigenfunctions of an integral operator defined in the paper.
}

\noindent
{\bfX Keywords :} {\rmX Inverse conductivity Problem, Nonlinear inverse problems, Electrical Impedance Tomography, Land mine detection.}
\medskip

\noindent {\bf 1. Introduction}

\noindent
It is well-known that the inverse conductivity problem is an
extremely ill-posed, non-linear inverse problem, and there has
been much interest in determining the class of conductivity
distributions that can be recovered from the boundary data, as
well as in the development of related reconstruction algorithms.
 The interest in this problem has been generated by
both  difficult theoretical challenges and by the important
medical, geophysical and industrial applications of this problem.
Much theoretical work has been related to the approach of  Calder\'on
 concerning the bijection
between the Neumann-to-Dirichlet operator, which relates  the
distribution of the injected currents  to the boundary values of
the induced electrical potential and the conductivity inside the
region \cite{calderon,kv85,sv88,nac95}.
The reconstruction procedures that have been proposed include a wide range of
iterative methods based on formulating the inverse problem as a
nonlinear optimisation problem. These techniques are quite demanding computationally particularly when addressing the three dimensional problem. This concern has encouraged the search for
reconstruction algorithms which reduce the computational demands
either by using some {\it a priori} information e.g.
\cite{fvog,brhanke,brhankevog} or by developing
non iterative procedures. Some of these methods \cite{brhanke, brhankevog} use a factorization approach while others are based on reformulating the inverse problems in terms of integral equations \cite{isaacson,andreea,cristiana}.

\noindent
In \cite{andreea}
we presented one such approach which we applied to a two  dimensional bounded domain while in
\cite{cristiana} we extended this work to an unbounded three dimensional
region. A feature of this method is the use of {\it a priori} information to find the regularized conductivity. This represents an additional requirement of our method but in some cases information on the internal structure of the object is already available. In return this enables us to make an estimate of the conductivity using a single data set. Another feature is that many of the calculations
involve working with analytical expressions containing the
eigenfunctions of the kernel of these equations, the computational
part being restricted of the introduction of the data, the
numerical evaluation of some of the analytic
 formul\ae \ and the solution of a final integral equation. However, this first version of our method
requires the knowledge on the boundary $\dOm$  of the region $\Om$ not
only of the electrical potential $\Phi$ and  its normal derivative
$\dpar\Phi/\dpar n$, but also of the electrical conductivity $\sigma$
and its normal derivative $\dpar\sigma /\dpar n$. In geophysics the conductivity function at the surface can
be measured by taking small soil samples, whilst in medical applications
this might be achieved using closely spaced electrodes. Although in
geophysics it is sometimes also possible to measure the derivative of the conductivity, in medical applications this is more difficult.

\noindent 
In this paper we describe a new version of
our approach which  no longer requires the knowledge of
$\dpar \sigma /\dpar n$, but uses only data which is easily
measurable on the boundary: namely the boundary values of the
conductivity function $\sigma$, the injected currents ( i.e. $\dpar\Phi /\dpar n |_{\dOm}$) and the corresponding
values of the induced potentials $\Phi|_{\dOm}$. We will show that this information is
sufficient to find a {\it regularised} solution
$\sigma_{reg}(\ve{x})$ for the conductivity. This new approach
differs from our previous work \cite{andreea,cristiana} also in
that the final step in the reconstruction algorithm involves the
solution of a first order linear partial differential equation
rather than one of second order.

\noindent
Throughout this paper we shall assume that the conductivity distribution
$\sigma\in C^1(\overline{\Om})$. Such functions are dense
in the larger class of the functions which have these properties
only piecewise and which contains many of the physically interesting
situations. In addition we shall assume that the domain $\Om$
is a bounded open set whose boundary $\dOm$ is sufficiently smooth,
namely of class $C^2$.

\noindent
Identification problems for elliptic equations have been the object of many studies in other fields as hydrology and diffusion equation \cite{yeh,richter, aless,chavent,lesnic}.

\medskip
 \noindent {\bf 2. Problem formulation}

\noindent From the conservation of  the current density $j(\ve{x})$ it
follows that in the case of a general isotropic
conductivity distribution  $\sigma(\ve{x})$, the
 potential $\Ph(\vx)$ satisfies :
\begin{equation}\label{invpr}
      \ve{\nabla} \cdot\left[ \, \sigma(\ve{x})\ve{\nabla}\Ph(\vx) \, \right]
      = 0 \, , \hskip 1cm
\ve{x} \in \Om \subset \Real^n \, , \quad n=2\,,3\, .
 \end{equation}
The inverse conductivity problem considered in this paper can be stated as follows:
from the values of  $\sigma$,  $\Ph$ and $\dpar \Ph / \dpar n$ on
the boundary $\dOm$, reconstruct the conductivity function $\sigma(\ve{x})$ in $ \Om$.

\noindent If we define   $\wt\sigma(\ve{x}) \id ln(\sigma(\ve{x}))$,
equation (\ref{invpr}) can be rewritten as
\begin{equation}\label{invpr2}
 \ve{\nabla}^2 \Ph(\vx)=-Y(\ve{x})\, ,\qqr
   \rm{ with} \qqr Y(\ve{x})\id \ve{\nabla}\wt\sigma(\ve{x})\cdot\ve{\nabla}
\Ph(\vx)\, .
 \end{equation}
This equation has the advantage that the ({\it not singular}) solution $\wt \sigma(\ve{x})$ to the inverse problem formulated in this way ensures the positivity of the conductivity $\sigma(\ve{x})$. To find $\wt\sigma(\ve{x})$ we will first
reconstruct the potential $\Ph(\vx)$ everywhere in $\Om$ from
the information available on the boundary. We will then solve the differential equation for  $\wt\sigma(\ve{x})$ using the method of characteristics. In this case, equation (\ref{invpr2}) reads
\begin{equation}\label{invpr3}
\frac{d\wt\sigma(\ve{x}(s))}{ds}=\frac{1}{|\ve{\nabla}\Phi|}Y(\ve{x}(s))\, ,
\end{equation}
where $s$ is the distance measured along the characteristics.

\noindent
By using Green's formula,  we can transform the partial differential equation
(\ref{invpr2}) and the boundary information into an equivalent pair of integral equations
\begin{eqnarray}
\Phi(\ve{x}) &=& \chi_D(\ve{x})  \, +  \, \int\limits_{\Om}  \,
d^n\ve{y} \, \GG_D(\ve{x},\ve{y}) \,
 Y(\ve{y}) \label{feq21}\\
\Phi(\ve{x})  &=&  \chi_N(\ve{x})
 \, +  \, \int\limits_{\Om}  \, d^n\ve{y} \, \GG_N(\ve{x},\ve{y}) \,
 Y(\ve{y}) \, ,\label{feq22}
\end{eqnarray}
where $\GG_D$ and $\GG_N$ are the Dirichlet and Neumann Green's
functions for Laplace's equation in $\Om$.
 Here $\chi_D$ and $\chi_N$ are the two  harmonic functions constructed
respectively from the measured Dirichlet and Neumann
boundary data, $\left.\Phi\right|_{\dOm}$ and
$\left. \dpar \Phi / \dpar n \right|_{\dOm}$ :
\begin{eqnarray} \chi_D(\ve{x})  &=& - \int\limits_{\dOm}
d^{n-1}\ve{z} \, {\dpar \GG_D \over \dpar n} (\ve{x},\ve{z}) \,
\left.\Phi(\ve{z})\right|_{\dOm} \\
\chi_N(\ve{x})\label{feq23}   &=&
\int\limits_{\dOm} d ^{n-1} \ve{z} \, \GG_N(\ve{x},\ve{z}) \,
\left. {{\dpar \Phi \over \dpar n}} \right|_{\dOm} \,+\,{C\over \mid
\dOm \mid} \ .\label{feq24}
\end{eqnarray}
where $C=\int\limits_{\dOm}\Phi(\ve{z})d^{n-1}\ve{z}$.
 These  functions are different unless
 $\sigma(\ve{x})$ is a constant.
\begin{theorem}\label{th1}
The knowledge of $\sigma(\ve{x})$ on $\dpar \Omega$ and of $\Phi(\vx)$ in $\Omega$
uniquely determines $\sigma(\ve{x})$ in $\Omega$.
\end{theorem}
\noindent {\bsl Proof} \hskip 0.3cm
Suppose that  $\wt\sigma_1$ and $\wt\sigma_2$ are two functions satisfying equation (\ref{invpr2}) and  $\wt\sigma_1\neq\wt\sigma_2$. Define $\wt\sigma_{diff}=\wt\sigma_1-\wt\sigma_2$. Then
\begin{equation}
 \nabla^2 \Phi(\ve{x})=-\nabla \wt\sigma_1(\ve{x})\cdot \nabla \Phi(\ve{x})=-\nabla \wt\sigma_2(\ve{x})\cdot \nabla \Phi(\ve{x})\,, \qqr \ve{x}\in \Omega
\end{equation}
\begin{equation}
\nabla\wt\sigma_{{\diff}}\cdot \ve{\nabla}\Phi(\vx) = 0\,,  \qqr \ve{x}\in \Omega \,.
\end{equation}
In other words, the function $\wt\sigma_{diff}(\ve{x})$ is constant
along the current lines inside $\Om$. Since $\sigma$ is known on $\dpar \Omega$ we have $\wt\sigma_{diff}(\ve{x})=0\,, \qqr \ve{x} \in \dpar\Omega$. Hence, provided that through any point $\ve{x} \in \Om$ there passes
a current line that intersects the boundary  $\dpar \Omega$ [see remark below and \cite{richter}], it follows that
\begin{equation}
\wt\sigma_{{\diff}}(\ve{x})= 0 \,, \qqr \ve{x}\in \Omega\,.
\end{equation}
Thus, $\wt\sigma$ is unique and it follows from the monotonicity of the logarithmic function that so too is $\sigma$. \hfill $\square$

\noindent {\bfX Remark} {\rmIX It is important to verify that through
each interior point of $\Om$ passes a current line which continues up to the
boundary. The danger with the current lines of a general vector field is that
they may spin indefinitely without reaching the boundary, as happens with
solenoidal fields. Since in our case the vector field is of gradient type, it is irrotational and so the field lines cannot
terminate inside the domain \cite{derham}
 but must continue up to the boundary.  Of course, there
might be some singular points where, for example, the gradient vanishes but as is  known  from Morse Theory \cite{milnor}, as long
we are  interested only in {\itIX generic cases}, these singular points are
isolated. There exist extreme situations (e.g. constant fields) for which the current lines avoid large regions, but
any generic  changes, even infinitesimal, will
lead to one of these ({\itIX generic}) {\itIX Morse Fields} in which $\sigma$ can be
constructed from its boundary values everywhere, except at most at some
isolated points where it can be found using continuity.}

\medskip
\noindent {\bf 3. The Direct Problem}

\noindent
Before considering  the Inverse Problem, let us see how one can
compute the potential $\Phi$ if the conductivity $\sigma$ is known. For instance, as shown in \cite{andreea,cristiana}, one may use the well known change of variable
$\tau=\sqrt{\sigma}$ to transform equation (\ref{invpr}) into
an integral equation for the function $\Psi\id \tau\Phi$. Another way
consists of
applying the operator $\ve{\nabla}_{\ve{x}}\wt\sigma(\ve{x})
\cdot \ve{\nabla}_{\ve{x}}$ to equation (\ref{feq21}) to give
\begin{equation}\label{feqY}
Y(\ve{x})= \ve{\nabla}_{\ve{x}}\wt\sigma(\ve{x}) \cdot
\ve{\nabla}_{\ve{x}}\chi_D(\ve{x})
 +  \, \int\limits_{\Om}  \, d^n\ve{y} \,K(\ve{x},\ve{y})
 \,
 Y(\ve{y})\, ,
\end{equation}
where $K(\ve{x},\ve{y})=\ve{\nabla}_{\ve{x}}\wt\sigma(\ve{x})\cdot
\ve{\nabla}_{\ve{x}}\GG_D(\ve{x},\ve{y})$. This is an integral
equation for the Laplacian $Y(\ve{x})$ of
 $\Phi(\ve{x})$.
Once $Y(\ve{x})$ is known, one can compute $\Phi(\ve{x})$  by means of a quadrature using the formula (\ref{feq21}) or (\ref{feq22}).

\noindent
It would be nice for the practical solution of equation
(\ref{feqY}) if the integral operator defined by the kernel
$K(\ve{x},\ve{y})$ were compact, since compact operators imply the
Fredholm Alternative \cite{hilbert}. For that it is sufficient that
$K\in L^2(\Om^2)$, but $ \int\limits_{\Om}  \, d^n\ve{x} \int\limits_{\Om}  \, d^n\ve{y} \, \left|
\ve{\nabla}_{\ve{x}}\wt\sigma(\ve{x})\cdot
\ve{\nabla}_{\ve{x}}\GG_D(\ve{x},\ve{y})\right|^2$ is divergent because of the singularity of
$\ve{\nabla}_{\ve{x}}\GG_D(\ve{x},\ve{y})$. However, we shall show that the situation can be
recovered by considering the first iteration of equation (\ref{feqY}) .
\begin{theorem}\label{th2}
If $\sigma\in C^1(\Om)$, the first iterated kernel of equation
(\ref{feqY}) $K_2(\ve{x},\ve{z})=
 \int\limits_{\Om}  \, d^n\ve{y}\,  K(\ve{x},\ve{y}) K(\ve{y},\ve{z})$
is Hilbert-Schmidt.
\end{theorem}
\noindent {\bsl Proof}  \hskip 0.3cm We shall prove that  the first iterated kernel
\begin{equation}\label{feqKit}
K_2 (\ve{x}, \ve{z}) \, \id \, \int\limits_{\Om} \,
d^3 \ve{y} \, \ve{\nabla}_{\ve{x}}\wt\sigma(\ve{x})\cdot
\ve{\nabla}_{\ve{x}}\GG_D(\ve{x},\ve{y})\ve{\nabla}_{\ve{y}}\wt\sigma(\ve{y})
\cdot \ve{\nabla}_{\ve{y}}\GG_D(\ve{y},\ve{z}) \,
\end{equation}
is  Hilbert-Schmidt. For smooth conductivities $\sigma$ in $\Omega$, any singular behaviour
in the integrand comes from the gradient of
 the Dirichlet Green's function. We consider here only the three dimensional case but the proof in two dimensions is similar. Although in two dimensions the first iterated kernel $K_2$ has a weak singularity (of logarithmic type) it is also Hilbert-Schmidt.
 
\noindent
We study only the leading singularity which is of the form
$$ \ve{\nabla}_{\ve{x}}\GG_D(\ve{x},\ve{y})\sim\frac{1}{4\pi}
  \frac{\ve{x}-\ve{y}}{|\ve{x}-\ve{y}|^3}\,. $$
Since $\wt\sigma\in C^1(\Omega)$, for any $\epsilon >0$ there exists a ball
$\Om_R\subset \Om$ of radius $R$ with centre $\ve{x}\in\Omega$  such that
$$|\ve{\nabla}_{\ve{x}}\wt\sigma(\ve{x})-
\ve{\nabla}_{\ve{y}}\wt\sigma(\ve{y})|<\epsilon\,, \qqr \qqr  \forall \ve{y}\in \Omega_R$$
Hence, in this ball we can approximate
$\ve{\nabla}_{\ve{y}}\wt\sigma(\ve{y})$ by
$\ve{\nabla}_{\ve{x}}\wt\sigma(\ve{x})$.
The most singular parts of
$K_2(\ve{x},\ve{z})$ can be written as
\begin{equation}\label{feqYit7}
K_2(\ve{x},\ve{z})\sim K_2^1(\ve{x},\ve{z})+K_2^2(\ve{x},\ve{z})
\end{equation}
where
$$K_2^1(\ve{x},\ve{z})=\frac{1}{16\pi^2}\int\limits_{\Om\backslash\Om_R} \,
d^3 \ve{y} \, \ve{\nabla}_{\ve{x}}\wt\sigma(\ve{x})\cdot
\frac{(\ve{x}-\ve{y})}{|\ve{x}-\ve{y}|^3}
\ve{\nabla}_{\ve{y}}\wt\sigma(\ve{y})
\cdot \frac{(\ve{y}-\ve{z})}{|\ve{y}-\ve{z}|^3}$$
and
$$K_2^2(\ve{x},\ve{z})=\frac{1}{16\pi^2}\int\limits_{\Om_R} \,
d^3 \ve{y} \, \ve{\nabla}_{\ve{x}}\wt\sigma(\ve{x})\cdot
\frac{(\ve{x}-\ve{y})}{|\ve{x}-\ve{y}|^3}
\ve{\nabla}_{\ve{x}}\wt\sigma(\ve{x})
\cdot \frac{(\ve{y}-\ve{z})}{|\ve{y}-\ve{z}|^3}\,.$$
\noindent Since the modulus of the gradient of $\wt\sigma$ is bounded on $\Om$,
we have
$$\left| K_2^1(\ve{x},\ve{z})\right|<\frac{Const}{16\pi^2}
\int\limits_{\Om\backslash\Om_R} \,d^3 y\,\,
 \frac{1}{|\ve{x}-\ve{y}|^2} \frac{1}{|\ve{y}-\ve{z}|^2}\, .
$$
When  studying the limit $r=|\ve{x}-\ve{z}| \rightarrow 0$ we
can take  $r<R/N<<R$ for some $N\in\Natural$.
Then, for $\ve{y}\in \Om\backslash\Om_R$, it follows that $|\ve{y}-\ve{x}|>R$
and $|\ve{y}-\ve{z}|>R(1-1/N)$, and hence
\begin{equation} \label{KK21}
\left| K_2^1(\ve{x},\ve{z})\right|<\frac{Const}{16\pi^2}
\frac{|\Om|}{(1-1/N)^2R^4}\, .
\end{equation}
\noindent To evaluate   $K_2^2(\ve{x},\ve{z})$ we define
$\ve{x}=(0,0,0)$, $\ve{z}=(0,0,r) $ and
$\ve{y}=(\rho\sin\theta\cos\phi,\rho\sin\theta\sin\phi,\rho\cos\theta)\,$,
where
$(\rho,\theta,\phi)\in \Real^3$ are local spherical coordinates in $\Om_R$.
The constant gradient of $\wt \sigma$ points in an arbitrary direction,
so for illustration we shall consider
$\ve{\nabla}_{\ve{x}}\wt\sigma(\ve{x})\,=\,(0,0,1)$. In this case
\begin{equation}\label{KK22}
K_2^2 (\ve{x}, \ve{z}) \, \sim \,\frac{1}{8\pi} \int\limits_{0}^R
d\rho\int\limits_{0}^{\pi}\sin\theta d\theta
\frac{\cos\theta(\rho\cos\theta-r)
}{\left(\rho^2-2\rho r\cos\theta+r^2\right)^{\frac{3}{2}}}\,
=\,-\frac{1}{12\pi R} \,.
\end{equation}
Combining equations (\ref{KK21}) and (\ref{KK22}) we see that
 $ K_2 (\ve{x}, \ve{z})$ remains finite as
 $|\ve{x}-\ve{z}|\id r \rightarrow 0$ and hence is
Hilbert-Schmidt.\hfill $\square$

\medskip
\noindent This theorem implies that the first iteration of equation
(\ref{feqY}), namely
\begin{equation}\label{feqYit}
Y(\ve{x})= \ve{\nabla}_{\ve{x}}\wt\sigma(\ve{x}) \cdot
\ve{\nabla}_{\ve{x}}\chi_D(\ve{x})
 +  \, \int\limits_{\Om}  \, d^n\ve{y} \,K(\ve{x},\ve{y})
\ve{\nabla}_{\ve{y}}\wt\sigma(\ve{y}) \cdot
\ve{\nabla}_{\ve{y}}\chi_D(\ve{y})
+  \, \int\limits_{\Om}\, d^n\ve{y}  \, K_2(\ve{x},\ve{y}))\, Y(\ve{y})\,
\end{equation}
can  be solved by means of the usual numerical procedures for
Fredholm equations. Once the function $Y(\ve{x})$ has been computed, the
potential $\Phi(\ve{x})$ can be obtained by means of the formul\ae \
(\ref{feq21}) or (\ref{feq22}). This procedure can then be used to perform simulations to study the performance of our method of
solving the Inverse Problem.

\medskip
\noindent
{\bf 4. The Unregularized Inverse Problem}

\noindent
Initially, our treatment of the inverse problem is similar to that
described in \cite{andreea,cristiana}. Subtracting equation (\ref{feq21}) from
(\ref{feq22}) we have
\begin{equation}\label{eqnr}
\chi(\ve{x}) -  \, \int\limits_{\Om} d^n\ve{y} \,
\KK (\ve{x},\ve{y}) \,  Y(\ve{y}) = 0
\end{equation}
where $$\KK (\ve{x},\ve{y}) =\
\GG_D(\ve{x},\ve{y})-\GG_N(\ve{x},\ve{y}) \qqr \qqr \rm{and} \qqr
\qqr \chi(\ve{x})=\chi_N(\ve{x})-\chi_D(\ve{x}) \, .$$
Equation(\ref{eqnr}) is an ill-posed Fredholm integral equation of
the first kind. Since the kernel $\KK(\ve{x},\ve{y})$ is the
difference of two Green's functions it is harmonic and symmetric.
The null space $N(\KK)$ of $\KK$ consists of all the functions
which are orthogonal to the harmonic functions in $\Om$. We can
prove the following result  concerning the potentials $\Phi_{\kk}$
related to this null space.
\begin{theorem}\label{th3}
Within the set of functions $\{\Phi_{\kk}\}$ that satisfy
$\nabla^2 \Phi_{\kk}= - Y_{\kk}$ where $Y_{\kk}\in N(\KK)$ there
exists at least one function such that
$\left.\Phi_{\kk}(\ve{x})\right|_{\dpar\Omega}=0$ and
$\left.\frac{\dpar\Phi_{\kk}}{\dpar
n}(\ve{x})\right|_{\dpar\Omega}=0$.
\end{theorem}
\noindent {\bf Note:} Such a function is, of course, completely
invisible to any measurement of potentials and currents on the
boundary.

\noindent {\bsl Proof } \hskip 0.3cm Since the Laplacian of any harmonic function is
identically zero, it is clear that the functions $\Phi_{\kk}$ are defined
only up to an harmonic function $\Phi_{h}$. So,
if the boundary values of some initial  $\Phi_{\kk}'$
are not zero, we can always find another function
$\Phi_{\kk}=\Phi_{\kk}'+\Phi_{h}$, having the same Laplacian $(-Y_{\kk})$,
 but with
\begin{equation}\label{chikborder}
\left.\Phi_{\kk}\right|_{\dpar\Om}=0\, .
\end{equation}
To this end it is sufficient to take
\begin{equation}
\Phi_h(\ve{x})=\int\limits_{\dpar\Om}d^{n-1}\ve{y}
\frac{\dpar\GG_D(\ve{x},\ve{y})}{\dpar n}\Phi_{\kk}'(\ve{y})\, ,
\end{equation}
where the function $\Phi_{h}$ is harmonic and has by construction
boundary values which are opposite to that of $\Phi_{\kk}'$. So
$\left.\Phi_{\kk}\right|_{\dpar\Om}=0$.

\noindent
To prove that the normal derivative of $\Phi_{\kk}$ is identically zero
on the boundary, we  use  Green's formula
\begin{equation}\label{green}
\int\limits_{\Om} \, d^{n}\ve{x} \,  \left( u(\ve{x})\, \nabla^2
\Phi_{\kk}(\ve{x}) \, - \,  \Phi_{\kk}(\ve{x})\nabla^2 u(\ve{x})
\right) \,= \, \int\limits_{\dOm} \, d^{n-1}\ve{x} \, \left( u(\ve{x})\,
{\dpar \Phi_{\kk}(\ve{x})\over\dpar n} \, - \,\Phi_{\kk}
(\ve{x})\, {\dpar u(\ve{x})\over\dpar n} \right) \, ,
\end{equation}
where $u(\ve{x})$ is any harmonic function in $\Omega$. Since 
$\nabla^2u=0$ and
since $\nabla^2 \Phi_{\kk}=-Y_{\kk}$ is orthogonal to any harmonic function,
the left hand side of the above equation is zero. But from (\ref{chikborder})
$\left.\Phi_{\kk}\right|_{\dpar\Om}\id 0$, we see that
\begin{equation}\label{proofth}
\int\limits_{\dOm} \, d^{n-1}\ve{x} \, u(\ve{x})\, {\dpar
\Phi_{\kk}(\ve{x})\over\dpar n} \,=\, 0 \, .
\end{equation}
Let us assume for a moment that at some given point $x_0\in \dpar\Omega$ the normal derivative were not zero,
$\left.\dpar \Phi_{\kk}(\ve{x}) / \dpar n
\right|_{\ve{x}=\ve{x_0}}\neq 0\, .$
Since according to our general assumptions the derivatives of
$\Phi$ are continuous in $\overline{\Om}$ --- i.e. also on $ \dOm$ ---
there exists a neighbourhood  on the boundary
$\dOm_{\epsilon,x_0} \subset \dOm$ of  $x_0\in \dpar \Om$,
on which this derivative has the same sign. Considering  for
$u(\ve{x})$ the harmonic measure defined by the boundary conditions
\begin{equation}
u(\ve{x})|_{\ve{x}\in\dOm}\,\,\,\,=
\,\,\,\left\{\begin{array}{r@{\quad,\quad}l}\displaystyle
   1& if\, \, \ve{x}\in \dOm_{\epsilon,x_0} \, , \\
   0& elsewhere\, \, on\, \, \dOm . \end
 {array}\right.
\end{equation}
we see that if $\dpar \Phi_{\kk}/\dpar n$ were not
identically  zero on $\dOm$,  the integral (\ref{proofth}) cannot be zero.
As a consequence both $\Phi_{\kk}(\ve{x})$ and
$\dpar \Phi_{\kk}(\ve{x})/\dpar n$ have to vanish identically on $\dOm$. 
\hfill $\square$

\medskip
\noindent
An example of such a function can be constructed explicitly from the function $Y_{\kk}$, as described in the following Lemma.
\begin{lemma}
The function $\Phi_{\kk}(\ve{x})= \int\limits_{\Om}  \,d^n\ve{y} \, 
\GG_D(\ve{x},\ve{y}) Y_{\kk}(\ve{y})$, where  $Y_{\kk}\in  N (\KK)$, has the properties stated in Theorem 3 i.e.:  $(i)\qqr \nabla^2 \Phi_{\kk}= - Y_{\kk}\,\qqr \ve{x}\in \Omega$, $(ii) \qqr \Phi_{\kk}(\ve{x})=0\,, \qqr \ve{x}\in \dpar \Omega$, $(iii)\qqr \frac{\dpar\Phi_{\kk}}{\dpar n}(\ve{x})=0\,,\qqr 
\ve{x}\in \dpar \Omega\,.$
\end{lemma}
\noindent {\bsl Proof } \hskip 0.3cm (i) If $x\in\Omega$ , then
$$\qqr \nabla^2\Phi_{\kk}(\ve{x})= \int\limits_{\Om}  \,d^n\ve{y} \, 
\nabla^2\GG_D(\ve{x},\ve{y})Y_{\kk}(\ve{y})=-\int\limits_{\Om}  
\,d^n\ve{y}\delta(\ve{x}-\ve{y})Y_{\kk}(\ve{y})=-Y_{\kk}(\ve{y})\,.$$
$(ii)$ If $\ve{x}\in\dpar\Omega$, recall that 
$\GG_D(\ve{x},\ve{y})=0$ and hence  $ \Phi_{\kk}(\ve{x})= \int\limits_{\Om}  
\,d^n\ve{y} \, \GG_D(\ve{x},\ve{y}) Y_{\kk}(\ve{y})=0\,.$

\noindent $(iii)$ 
If $Y_{\kk}\in  N (\KK)$ and $\KK(\ve{x},\ve{y})=\GG_D(\ve{x},\ve{y})-\GG_N(\ve{x},\ve{y})$ it follows that 
$$\Phi_{\kk}(\ve{x})= \int\limits_{\Om}  \,d^n\ve{y} \, \GG_N(\ve{x},\ve{y}) Y_{\kk}(\ve{y})\, , \qqr \ve{x}\in\Omega\,$$
is also a valid representation for $\Phi_{\kk}$.
Moreover,  $\left.\frac{\dpar \GG_N(\ve{x},\ve{y})}{\dpar n}\right|_{\ve{x}\in\dpar\Omega} =-{{1}\over {|\dOm |}}\,$, so that
$$\left.\frac{\dpar\Phi_{\kk}}{\dpar n}(\ve{x})\right|_{\ve{x}\in \dpar\Omega}=-{{1}\over {|\dOm |}}\int\limits_{\Om}  \,d^n\ve{y} \,Y_{\kk}(\ve{y})=0\,$$
since $Y_{\kk}(\ve{y})$ is orthogonal to any harmonic
function in $\Om$ and the constant $1$ is an harmonic function.
\hfill $\square$

\medskip
\noindent
Explicit analytical expressions of such functions $Y_{\kk}\in N(\KK)$ and their respective $\Phi_{\kk}$ for the unit disk are given in Section 6. 

\noindent
At this stage an apparent paradox may emerge. Consider the case $\Phi\id\Phi_{\kk}$ corresponding to $Y_{\kk}\in N(\KK)$. Since $\ve{\nabla}\Phi_{\kk}$ is in general  not identically zero throughout $\Omega$ (see, for example, the explicit expressions given in Section 6), one may be tempted to reconstruct $\wt\sigma$ inside $\Omega$ by solving along the characteristics the differential equation  (\ref{invpr3}) using the given boundary data $\wt\sigma(\ve{x})=\ln (\sigma(\ve{x}))$ for $\ve{x}\in \dpar \Omega$. However,
\begin{equation}\label{energy}
\int\limits_{\Om} \, d^{n}\ve{x} \, \sigma(\ve{x})\,
 \mid\nabla \Phi(\ve{x})\mid^2 \,
 \,= \, \int\limits_{\dOm} \, d^{n-1} \ve{x} \, \sigma(\ve{x})\,\Phi_{\kk}(\ve{x})\,{\dpar \Phi_{\kk}(\ve{x})\over\dpar n} \,=0 \,,
\end{equation}
since both $\left.\Phi_{\kk}(\ve{x})\right|_{\dpar\Omega}$ and $\left.\frac{\dpar\Phi_{\kk}}{\dpar n}(\ve{x})\right|_{\dpar\Omega}$ are identically zero. From the non-negativity of $\sigma(\ve{x})=\exp(\wt\sigma(\ve{x}))$, the left hand side is positively defined and so $\sigma$ has to be zero everywhere where $\ve{\nabla}\Phi_{\kk}$ is not. The key of this apparent dilema is that equation (\ref{invpr3}) cannot be solved since the normal component of $\ve{\nabla}\Phi_{\kk}$ vanishes when $\ve{x}$ approaches the boundary. This represents a barrier that cannot be overpassed ($\left|\frac{d\wt\sigma}{ds}\right|\rightarrow \infty$) by the information conveyed by the equation (\ref{invpr3}). On the other hand, if $\sigma$ is constant, from equation (\ref{invpr2}) it follows that $Y_{\kk}$ has to vanish identically throughout $\Omega$, i.e. the coefficients $A_{k,n}$ and $B_{k,n}$ in expression (\ref{yk}) have to be zero.

\medskip
\noindent {\bf 5. The Regularized Integral Equation}

\begin{figure}[ht]
\centerline{\epsfig{file=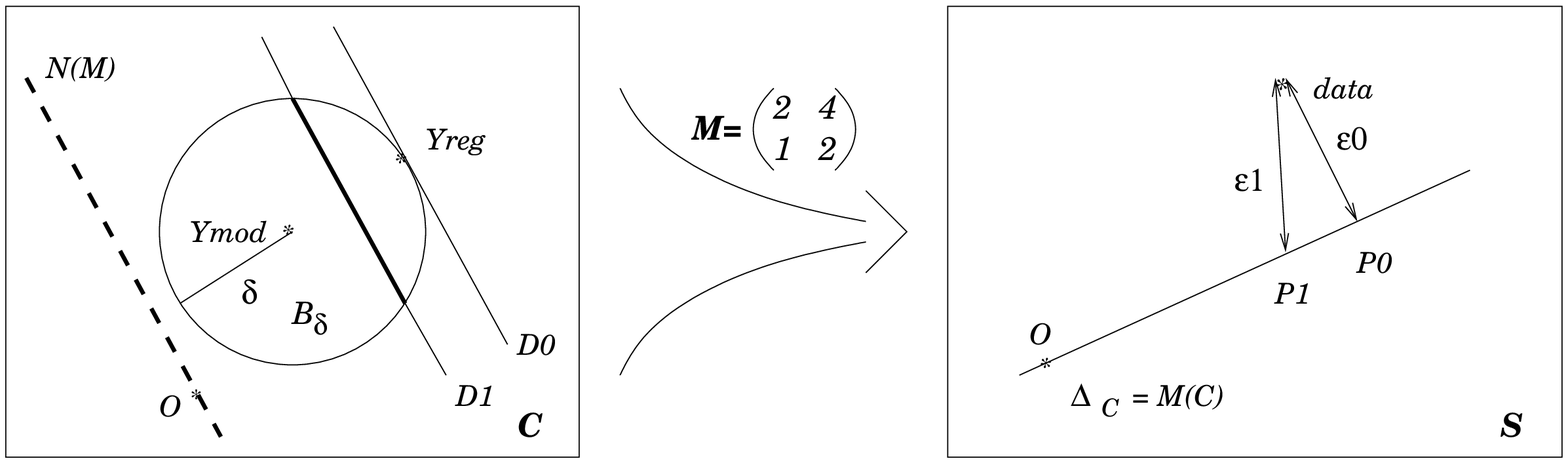,height=3cm,width=12cm}}
\caption{}
\hskip 0.5cm{\begin{minipage}{15cm}
\footnotesize For the Tikhonov regularisation it is not necessary that the 
mapping $\MM$ \  between the {\it parameter} space $\CC$ 
and the {\it data}  space $\SS$ should have a true inverse. 
The existence of a generalized inverse $\widehat{\MM^{-1}}$ is 
sufficient. In the above figure the mapping transforming 
the $\CC$ plane into the straight line $\Delta_C \in \SS$ is provided 
by the non-invertible matrix $\pmatrix{2&4\cr 1&2\cr}$ while the generalized
inverse $\widehat{\MM^{-1}}$ transforms the points $P_1,\ P_0 \in \Delta_C$
into the lines $D_1,\ D_0 \in \CC$.

These low dimensional graphs may be misleading, as in infinite
dimensions it might happen that there does not exist a smallest
distance $\epsilon_0$ between the {\it admissible} set $\Delta_C=\MM(\CC)$ and the {\it data} $\notin\Delta_C$. Therefore, in infinite dimensional problems it is preferable to consider images of balls $\BB$ from $\CC$ rather than 
$\MM(\CC)$ itself. Since in most physical problems the theorem of Banach-Alaoglu \cite{barry,yosida} also applies \cite{mihaela} , the images $\MM(\BB)$ are true compact sets and the minimal distance $\epsilon_0$ of such a set to the {\it data} is well defined. Nevertheless,
the main features of these problems are well depicted by these figures, where the interesting
points on  the straight lines $ D_1\, ..., D_0$  are their
 intersections with the ball $\BB_{\delta}$ of radius $\delta$ around $Y_{mod}$. Consequently, in our case the solution of the extremal problem will be the point $Y_{reg}\in D_0$. 
\end{minipage}}
\label{regfig}
\end{figure}

\noindent
Equation (\ref{eqnr}) is a Fredholm equation of the first kind
which is  notoriously ill-posed and whose
solution is extremely unstable with respect to the data
measured on the boundary $\dOm$. We shall stabilize it by using Tikhonov regularization and a model
function $Y_{mod}(\ve{x})$ as a first estimate of the solution (see Figure \ref{regfig}). This function
can be obtained starting with a model conductivity $\sigma_{mod}$ and solving
equation (\ref{feqY}), or better its first iterate (\ref{feqYit}), to find
$Y_{mod}(\ve{x})$. The function $\sigma_{mod}$ can be obtained in a variety of ways including the use of prior knowledge from other imaging modalities, previous results in time evolving situations or using simple reconstruction estimates using fast algorithms. One example of the latter is a Factorization method described in \cite{brhanke, brhankevog}. In this method a large number of current patterns are applied to the object and using this data a simple test at each point within $\Omega$ is used to determine a piecewise constant internal structure. This algorithm gives good results in many circumstances although practical numerical difficulties introduce instability. Nevertheless, we can use a continuous approximation to these results as a suitable $\sigma_{mod}$. 

\noindent
The use of the regularizing function $Y_{mod}$ leads to
a Fredholm equation of the second kind which can be solved explicitly and
which is stable with respect to the errors in the input data. Moreover the
null space of $\KK$ plays no particular role in the solution of Fredholm
equation  of the second kind which is unique unless the Lagrange multiplier
defined below happens to be an eigenvalue of $\KK$.

\noindent
Indeed, the input data and hence the function $\chi(\ve{x})$ always contain some errors. Consequently we no longer require that equation (\ref{eqnr})
 should be satisfied
exactly, but we ask instead that the $L^2$ norm of the left hand side of
(\ref{eqnr}) should be small while the solution  $Y_{reg}$ should be close
to a model function  $ Y_{mod}$. Hence our problem reduces to  finding
\begin{equation}\label{delta}
\epsilon_0 =
min \, \|\chi (\ve{x}) \, - \,{\bf K}[Y_{reg}] (\ve{x}) \, \|_{L^2} \, ,
  \qqr {\rm subject} \qqr {\rm to}\qqr \left\|Y_{reg}(\ve{x}) \,
 - \, Y_{mod}  (\ve{x})\right\|_{L^2}\leq \delta\, .
\end{equation}
Here ${\bf K}[Y_{reg}] (\ve{x})\id\int \, d^n\ve{y} \,
 \KK(\ve{x},\ve{y}) \,Y_{reg} (\ve{y})$. In practical terms this means
that instead of eq.(\ref{eqnr}) we consider the Lagrange
multiplier formulation of Tikhonov regularization \cite{engl,sorin},
namely we seek the unrestricted minimum of the functional
$$\|\chi (\ve{x}) \, - {\bf K}[Y_{reg}] (\ve{x}) \|_{L^2}^2\,+ \, \mu \,
\left( \left\|Y_{reg}(\ve{x}) \, - \, Y_{mod}
 (\ve{x})\right\|_{L^2}^2 \,-\,\delta^2\right) \, ,$$
where $\mu$ is a Lagrange multiplier. 

\noindent
In many cases of practical interest
the target function depends also on a small number  of
parameters $\alpha$, for instance connected with the translation and/or
rotation Euclidean group. Consequently we could include these parameters in
the model function  $Y_{mod}(\ve{x};\alpha)$ and seek the minimum
$\epsilon_{0,min} \id \epsilon_0(\alpha_0)$ with respect to these parameters. Such an example is discussed in Section 8. Here $\epsilon_0$ which is a functional of the data, or its minimum  $\epsilon_{0,min}$ with respect to the possible
parameters $\alpha$, measures the degree \cite{sorin} to which the initial
 integral equation (\ref{eqnr}) is satisfied. 
In other words, it reflects the accuracy of the experimental
data by means of which (see eq.(\ref{feq23})) the input function
$\chi (\ve{x})$  has been constructed.

\noindent The minimisation process  yields the following
regularized integral equation:
\begin{equation}\label{eqreg}
Y_{reg}(\ve{x}) \, = \, Y_{mod} (\ve{x}) \, + \, \ll \,
\int\limits_{\Om} \, d^n \ve{y} \, \KK (\ve{x},\ve{y}) \, \chi (\ve{y})
\, -  \, \ll \,
\int\limits_{\Om} \, d^n \ve{y} \,
 \KK_2 (\ve{x}, \ve{y}) \,  Y_{reg}(\ve{y})
\end{equation}
\noindent
where $\displaystyle \KK_2 (\ve{x}, \ve{y}) \, \id \, \int\limits_{\Om} \,
d^n \ve{z} \, \KK (\ve{x},\ve{z}) \, \KK (\ve{z},\ve{y}) \, $ and
$\ll \id  1/\mm$.

\noindent
The eigenfunctions $\left\{u_k\right\}$  and eigenvalues $\left\{\lambda_k\right\}$ of  $\KK(\ve{x}, \ve{y})$ satisfy $u_k(\ve{x})=\lambda_k\int\limits_{\Om} \,d^n \ve{y} \, \KK (\ve{x},\ve{y}) \,u_k(\ve{y})$. If the kernel $\KK (\ve{x}, \ve{y})$ can be expressed in the form
$$\KK(\ve{x}, \ve{y}) \, = \, \sum_{k=1}^{\infty} { u_k (\ve{x}) \, u_k
(\ve{y}) \over \ll_k }\, ,$$
we have a straightforward way to compute the eigenfunctions and
eigenvalues and we can expand the iterated kernel $\KK_2$ as:
$$\KK_2 (\ve{x}, \ve{y}) \, = \, \sum_{k=1}^{\infty} { u_k (\ve{x})
\, u_k (\ve{y})
 \over \ll_k^2 }\, .$$
This is illustrated in the two dimensional example discussed
in Section 6 (see also \cite{cris2}).

\noindent Projecting Eq.(\ref{eqreg})  onto $u_k$ we obtain
$$ Y_{reg,k} \, = \,  Y_{mod,k} \,  + \, {\ll \over \ll_k} \,  \chi_k  \, -
 \,   {\ll \over \ll_k^2} \, Y_{reg,k}$$ where
$$ Y_{reg,k} = \int\limits_{\Om} d^n \ve{x} \, Y_{reg}(\ve{x})
u_k (\ve{x}) \;
, \; Y_{mod,k}  = \int\limits_{\Om} d^n  \ve{x} \, Y_{mod} (\ve{x})
u_k (\ve{x}) \; , \;
\chi_k = \int\limits_{\Om} d^n \ve{x}  \, \chi (\ve{x})
u_k (\ve{x}) \, ,$$
and hence the following explicit expression for the
solution $Y_{reg}(\ve{x})$:
\begin{equation} \label{Yreg}
Y_{reg}(\ve{x}) \, = \, Y_{mod} (\ve{x}) \, + \, \ll \,
\int\limits_{\Om} \, d\ve{y}^n \, \KK (\ve{x},\ve{y}) \, \chi (\ve{y}) \,
-  \, \ll \, \sum_{k=1}^{\infty} \,{\displaystyle
Y_{mod,k} \,  + \,  {\ll \over \ll_k} \,  \chi_k \over
\ll \, + \, \ll_k^2 } \,  u_k (\ve{x})\, .
\end{equation}
Once $Y_{reg}(\ve{x})$ is known, the potential $  \Phi_{reg}(\ve{x})$
can be obtained by means of the integral
\begin{equation}
 \Phi_{reg}(\ve{x}) \, = \,
\chi_D(\ve{x})  \, +  \, \int\limits_{\Om}  \, d^n\ve{y} \,
\GG_D(\ve{x},\ve{y}) \, Y_{reg}(\ve{y})\, .
\end{equation}
Finally, to determine the regularized $\sigma_{reg}(\ve{x})$,
we can use the method of characteristics to solve the first order partial
differential equation:
\begin{equation}\label{pde}
\nabla\wt\sigma_{reg}(\ve{x}) \cdot \nabla\Phi_{reg}(\ve{x})-
Y_{reg}(\ve{x})=0 \,,  \qqr \ve{x}\in\Om
\end{equation}
subject to the known boundary values
$\left.\wt\sigma(\ve{x})\right|_{\ve{x}\in\dpar\Om}$.

\medskip
\noindent {\bf 6. A two dimensional example: the unit disk}

\noindent
In this case the relevant Green's functions are:
\begin{equation}\label{greenD2}
\GG_D(r,\theta;\rho,\vartheta) \, = \,  - \, {1 \over 4 \pp} \, \log
{r^2 + \rho^2 - 2 r \rho \cos \, (\th -\vartheta) \over
 1 + r^2 \rho^2 - 2 r \rho \cos \, (\th -\vartheta)}
\end{equation}
\begin{equation}\label{greenN2}
\GG_N(r,\theta;\rho,\vartheta) \, = \, - \, {1 \over 4 \pp} \, \log
\left( (r^2 + \rho^2 - 2 r \rho \cos \, (\th -\vartheta)) \cdot (1 + r^2
\rho^2 - 2 r \rho \cos \, (\th -\vartheta)) \right) \,
\end{equation}
\noindent so that
\begin{equation}
\KK(r,\th;\rho,\vartheta)={1 \over 2 \pp} \, \log \,
(1+r^2 \rho^2 - 2 r \rho \cos \, (\th - \vartheta))\,\displaystyle - \,
\sum_{k=1}^{\infty}{r^k \rho^k \cos(k(\th-\vartheta)) \over \pp k} \,.
\end{equation}
As remarked in Section 5, this series expansion enables us to find  the eigenfunctions
and  eigenvalues of $\ve{\KK}$.
Indeed, the functions $\displaystyle u_{k}^1 (r,\th)
=   \sqrt{(2 k + 2) / \pp} \, \, r^k \cos \, k \th$ and
$\displaystyle u_{k}^2 (r,\th) = \sqrt{(2 k + 2) /\pp} \, \, r^k
\sin \, k \th$ are orthonormal on the unit disk, so that $\KK$ can
be rewritten as $$\KK (r,\th; \rho,\vartheta) \, = \, \sum_{k=1}^{\infty}
\sum_{j=1}^{2} \, {u_k^j \, (r,\th) \, u_k^j \, (\rho,\vartheta) \over
\ll_k} \, ,$$ \noindent with $\ll_k = -2 k (k+1)$.

\noindent
From expansion (\ref{Yreg}) we obtain the following explicit expression for
the solution of the regularized equation
\begin{equation}\label{Yreg2}
Y_{reg}(r,\th) \, = \, Y_{mod} (r,\th) \, + \, \ll \,
\int_0^1 \rho d\rho \int_0^{2\pi}\! d\vartheta \, \KK (r,\th;\rho,\vartheta) \,
\chi(\rho,\vartheta) \, -  \, \ll \, \sum_{k=1}^{\infty} \, \sum_{j=1}^{2}
{\displaystyle Y_{mod,k}^j \,  - \,  {\ll \over 2 k (k+1)} \,
\chi_k^j \over \ll \, + \, 4 k^2 (k+1)^2 } \,  u_k^j (r,\theta).
\end{equation}
\noindent To find $\sigma_{reg}$ we proceed in the way discussed at the
end of Section 5.
\begin{lemma}
For the unit disk the functions $Y_{\kk}\in N(\KK)$ have the explicit form
\begin{equation}\label{yk}
Y_{\kk}(r,\th)=\displaystyle\sum_{k=1}^\infty\sum_{n=1}^\infty G_n(k+2,k+2,r)
\Bigl(A_{k,n}\cos(k\th)+B_{k,n}\sin(k\th)\Bigr)
\,,
\end{equation}
where $A_{k,n}$ and $B_{k,n}$ are arbitrary constants.
\end{lemma}
\noindent {\bsl Proof} \hskip 0.3cm If $Y_{\kk}\in N(\KK)$, then
$$\int_0^1 \rho d\rho\int_0^{2\pi}d\vartheta\KK (r,\th;\rho,\vartheta)Y_{\kk}(\rho,\vartheta)= \sum_{k=1}^{\infty}
\sum_{j=1}^{2} \, {u_k^j \, (r,\th) \over
\ll_k}\int_0^1 \rho d\rho\int_0^{2\pi}d\vartheta\,u_k^j(\rho,\vartheta)Y_{\kk}(\rho,\vartheta)=0\,.$$
We start with a function $Y_{\kk}$ of the form
$$Y_{\kk}(r,\th)=\Re_k(r)\cos(k\th)\,,\qqr k=1,2,3,\ldots\,.$$
which, in order to be in $N(\KK)$ must satisfy
 $$\int_0^1 \rho d\rho\int_0^{2\pi}d\vartheta u_k^1(r,\th;
\rho,\vartheta)Y_{\kL}(\rho,\vartheta)=-\sqrt{(2k+2)\pi}
\int_0^1 d\rho \ \rho^{k+1}\Re_k(\rho)=0\,.$$
For this it is sufficient that
$$\Re_k(r,n)=G_n(k+2,k+2,r)\,, {\rm for }\ n\geq 1\,,$$ where
$G_n(k+2,k+2,r)$ are the Jacobi polynomials of order $n$ \cite{hilbert}.

\noindent If we consider
$$Y_{\kk}(r,\th)=\Re_k(r)\sin(k \th)\,,\qqr k=1,2,3,\ldots\,.$$
 we find a similar result. Our theorem follows by taking linear combinations of these results. \hfill $\square$

\noindent
Example: If we consider $Y_{\kk}(r,\th)= G_1(3,3,r)\cos(\th)=\Bigl(1-\frac{4}{3}r\Bigr)\cos(\th)\,\in N(\KK)$ we can compute directly 
$$ \Phi_{\kk}(r,\th)= \int_{0}^1\rho d\rho \int_{0}^{2 \pi} d \vartheta 
\GG_D(r, \theta ; \rho, \vartheta) Y_{\kk}(\rho \vartheta)=\frac{1}{6}r(r-1)^2\cos(\th)\,.$$

\medskip

\noindent {\bf 7.  A numerical example}

\begin{figure}[ht]
\centering
\mbox{\hskip -2cm\subfigure[]{\epsfig{file=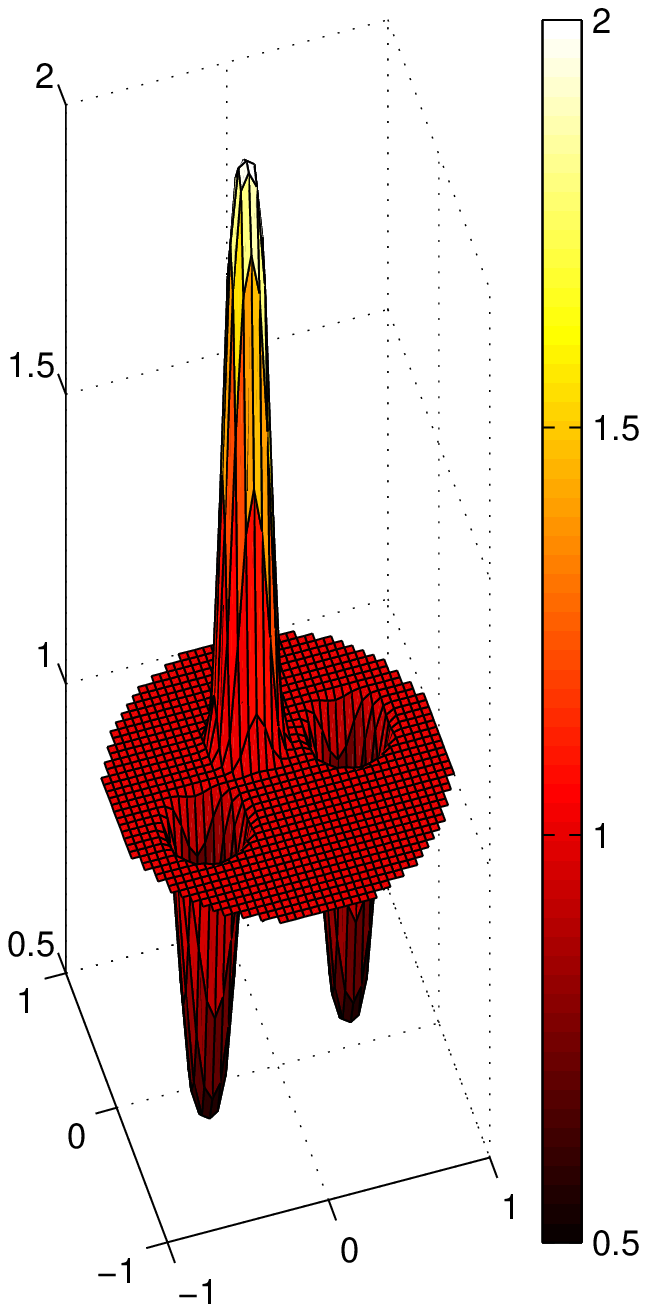,height=5cm,width=8cm}\label{exfig}} \hskip 1cm
\subfigure[]{\epsfig{file=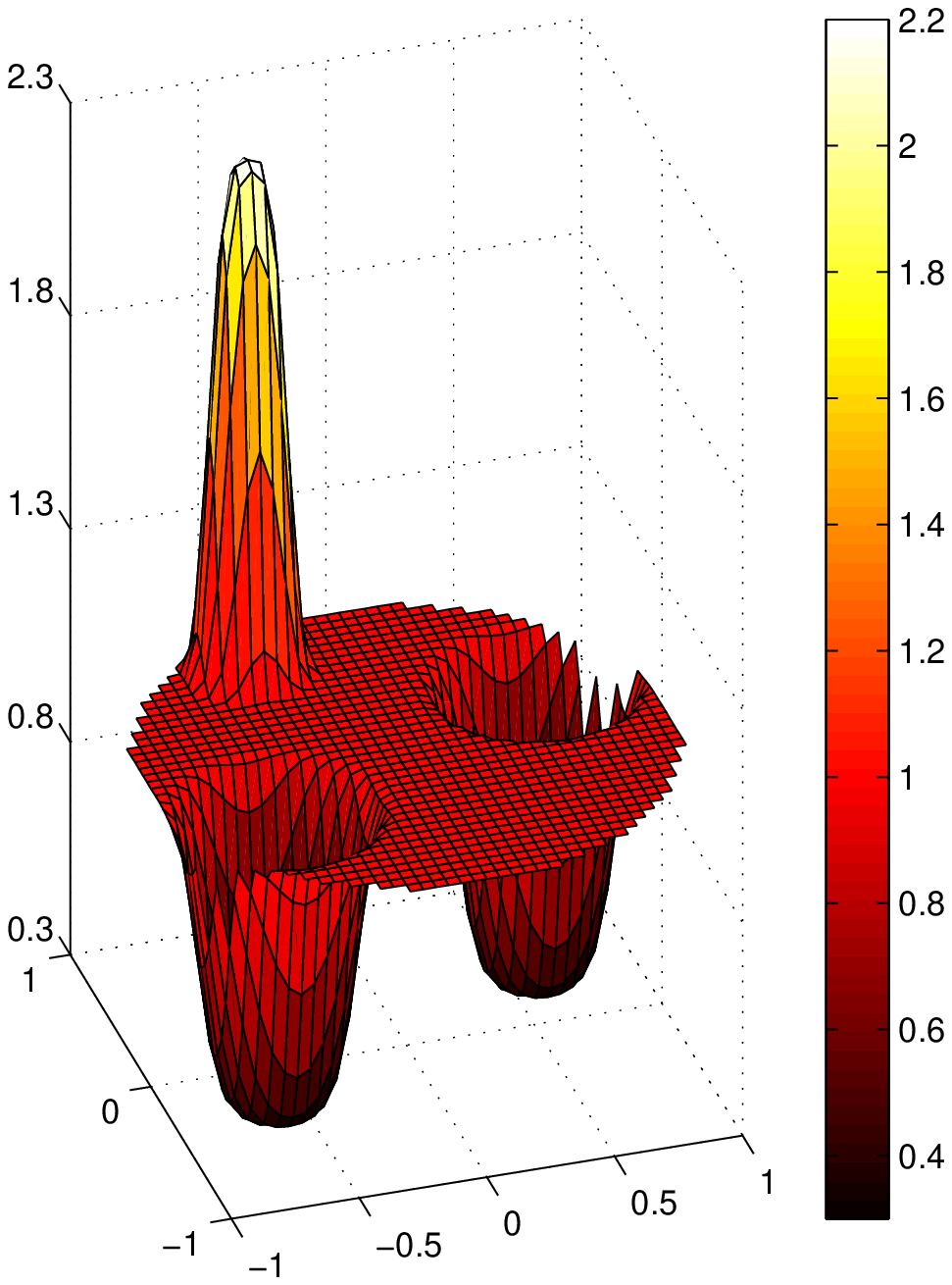,height=5cm,width=6cm}\label{modfig}}}
\caption{\rmX (a) The exact conductivity $\sigma_{exact}$; \qqr \qqr (b) The model conductivity $\sigma_{mod}$.}
\end{figure}

\noindent
We present here a numerical example to illustrate the performance of our algorithm. In this example we attempt to reconstruct a conductivity
distribution consisting of two low conductivity regions and a high
conductivity region within an uniform background. We represent this
distribution mathematically by a function of the form:
\begin{eqnarray}
\sigma_{exact}(x,y)&=&1-0.5\exp(-a((x-0.5)^2+(y-0.2)^2)^2)+\exp(-b((x+0.1)^2+(y-0.3)^2)^2)
\nonumber \\&-&0.5\exp(-a((x+0.5)^2+(y+0.2)^2)^2)\,,\qqr
x^2+y^2\leq 1\, ,
\end{eqnarray}
which has a  maximum at $(x,y)=(-0.1,0.3)$ and two minima
at$(x,y)=(-0.5,-0.2)$ and $(x,y)=(0.5,0.2)$ (see Figure \ref{exfig}). We
chose $a=1000$ and $b=1500$.

\noindent
We shall consider the input currents
$$\left.\sigma(1 ,\theta)\frac{\dpar
\Phi}{\dpar n}\right|_{\dOm}\!(\theta)=\sin(m\theta)\,,\qqr m=1,2,\ldots\,.$$
To simulate the measured values of the potential on the boundary we
have to solve first the direct problem. The detailed numerical simulations can be found in \cite{cristh}. With these data
we can compute then the  input function $\chi$  of our integral equation
$$\chi(r,\theta)=\displaystyle\int_0^{2 \pi}d\vartheta
\left\{\GG_N(r,\th;1,\vartheta)\left.\frac{\dpar \Phi}{\dpar
n}\right|_{\dOm}\!(\vartheta)+\left[\frac{1}{2\pi}+\frac{\dpar \GG_D}
{\dpar n}(r,\theta;1,\vartheta)\right]
\Phi(1,\vartheta)\right\}\,.$$
To calculate the regularized solution $Y_{reg}$ we need a model
function $Y_{mod}$. To achieve this we need a model function for the
conductivity $\sigma_{mod}$, which is in fact
our {\it a priori} information. For the model function $\sigma_{mod}$ we have used a conductivity
distribution in which

\noindent - the positions of the low conductivity regions are quite
well known but the size of these regions is more uncertain

\noindent - the size of the high conductivity region is quite well
known but the position of this region is uncertain.

\noindent We have represented this model function mathematically
as:
\begin{eqnarray}
\sigma_{mod}(x,y)&=&1-0.7\exp(-a((x-0.6)^2+(y-0.3)^2)^3)+1.2\exp(-b((x+0.4)^2+
(y-0.7)^2)^2) \nonumber
\\&-&0.7\exp(-a((x+0.6)^2+(y+0.3)^2)^3)\,,
\end{eqnarray}
which lies at a distance
$d\id\|\sigma_{mod}-\sigma_{exact}\|_{L^2}=0.29\|\sigma_{exact}\|_{L^2}$
from the true function (see Figure \ref{modfig}). This type of function could be a smooth approximation to piecewise constant approximates produced by \cite{ brhanke, brhankevog} as discussed in Section 5. 
By solving the forward problem as described in Section 3, we find the
model potential $\Phi_{mod}$ and hence the model functions
$Y_{mod}(x,y)=\nabla\wt\sigma_{mod}(x,y)\cdot\nabla\Phi_{mod}(x,y)$.
\noindent
By solving the forward problem as described in Section 3, we find the
model potential $\Phi_{mod}$ and hence the model functions
$Y_{mod}(x,y)=\nabla\wt\sigma_{mod}(x,y)\cdot\nabla\Phi_{mod}(x,y)$.
The integral which appears in (\ref{Yreg2}) can be computed
numerically: $$\int_0^1 \rho d\rho\int_0^{2\pi}d\vartheta
\KK(r,\th;\rho,\vartheta)\chi(\rho,\vartheta)=\sum_{l=1}^{172}w_l
\KK(r,\th;r_l,\th_l)\chi(r_l,\th_l)\,,$$ where $\{w_l;r_l,\th_l\}$
is a set of quadrature weights and points for the unit disk
\cite{engl2}. Since, the infinite series which appear in equation
(\ref{Yreg2}) converges rapidly it was sufficient to consider only
fifty terms to reach a precision of $10^{-8}$.

\noindent
There are many ways of determining the value of the parameter
$\lambda$ \cite{engl,hansen} and we have used the constraint (\ref{delta}) for a given value of $\delta$.
 Once $Y_{reg}$ is known, the calculation of the potential $\Phi$ inside
 the unit disk is straightforward:
$$
\Phi(r,\th)=\chi_D(r,\th)+\sum_{l=1}^{172}w_l\GG_D(r,\th;r_l,\th_l)
Y_{reg}(r_l,\th_l)\,,$$
where
$$\chi_D(r,\th)=\frac{1-r^2}{2\pi}\int_0^{2\pi}\frac{\Phi(1,\vartheta)}
{1+r^2-2r\cos(\th-\vartheta)}d\vartheta\,.$$
The final step is to solve our first order partial differential
equation (\ref{pde}). The method of characteristics transforms this equation
into an equivalent system of ordinary differential equations which can be
written in cartesian coordinates as:
\begin{equation}
\frac{dx}{ds}  = \frac{d\Phi}{dx}(x(s),y(s)) \, , \qqr
\frac{dy}{ds} = \frac{d\Phi}{dy}(x(s),y(s))\,,\qqr
\frac{d\wt\sigma}{ds} = Y_{reg}(x(s),y(s))\, .
\end{equation}
We have used sixty characteristic curves, the initial conditions for the
$i$-th characteristic being:
\begin{equation}
x_0^i=\cos\th_0^i\,,\qqr  y_0^i=\sin\th_0^i\, ,\qqr
 \theta_0^i=2\pi/i\, ,\ for\  i=1\, ...\,60\, ,
\end{equation}
\begin{eqnarray*}
\wt\sigma_0^i&=&\ln(1-0.5\exp(-1000(1.29-\cos\th_0^i-0.4\sin\th_0^i)^2)+\exp(-1500(1.1+0.2\cos\th_0^i-0.6\sin\th_0^i)^2)
\\&-& 0.5\exp(-1000(1.29+\cos\th_0^i+0.4\sin\th_0)^2))\,.
\end{eqnarray*}

\begin{figure}[H]
\centerline{\epsfig{file=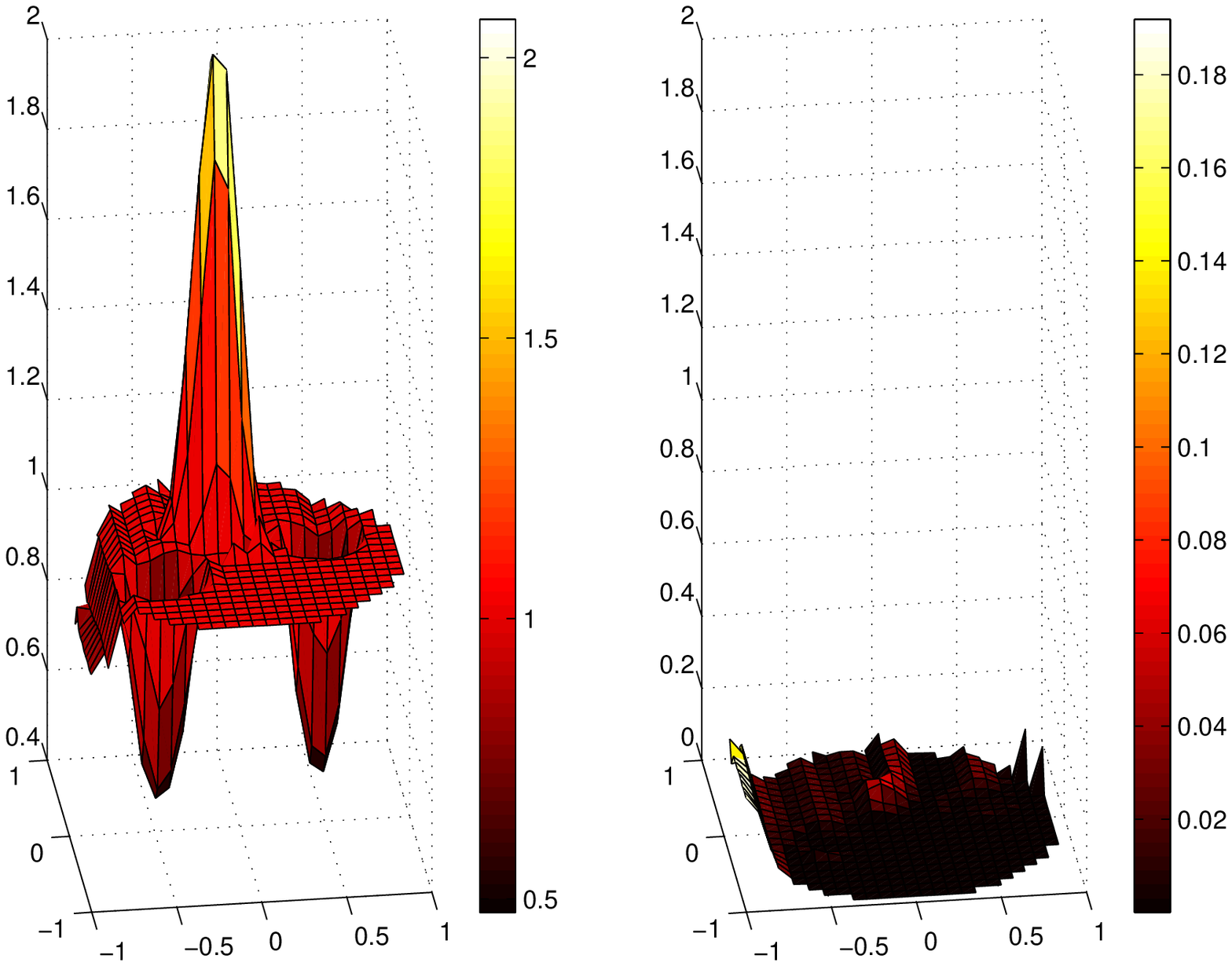, height=5cm,width=14cm}}
\centerline{(a)\hskip 6cm (b)}
\caption{\rmX The reconstructed conductivity using
$\sin(\th)$ for data with random 2\% errors}
\centerline{\rmX (a) - the reconstructed conductivity $\sigma(x,y)$; \qqr \qqr (b) - the absolute errors
$\mid\sigma(x,y)-\sigma_{exact}(x,y)\mid$ }
\label{sig1fig}
\end{figure}

\begin{figure}[H]
\centerline{\epsfig{file=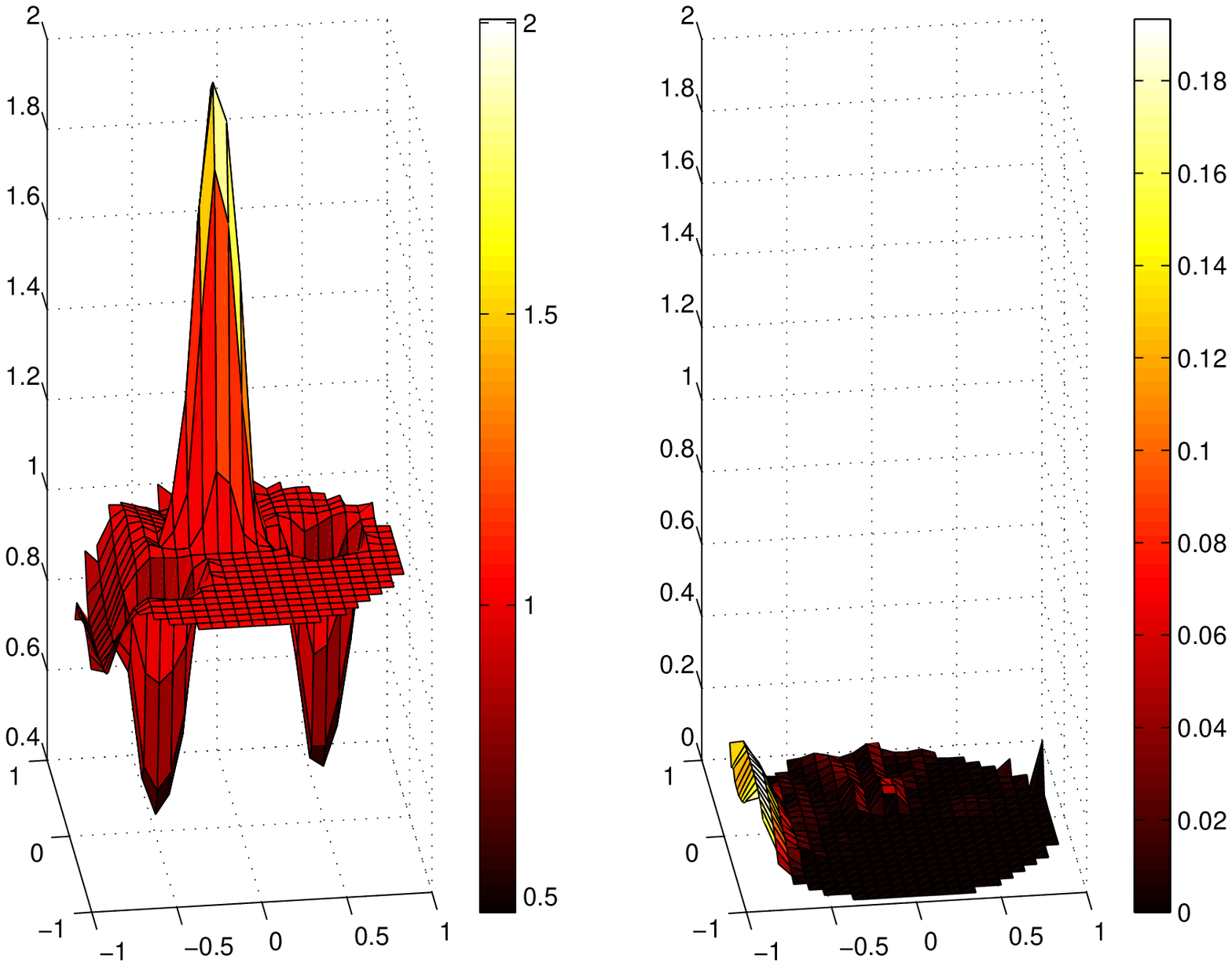, height=5cm,width=14cm}}
\centerline{(a)\hskip 6cm (b)}
\caption{\rmX The reconstructed conductivity using
$\sin(2\th)$ for data with random 2\% errors}
\centerline{\rmX (a) - the reconstructed conductivity $\sigma(x,y)$; \qqr \qqr (b) - the absolute errors
$\mid\sigma(x,y)-\sigma_{exact}(x,y)\mid$ }
\label{sig2fig}
\end{figure}

\noindent
We present in Figures \ref{sig1fig} and \ref{sig2fig} the results obtained for two current 
patterns with $m=1$ and $2$ for data with 2\% random
errors. The reconstructed conductivity obtained
 lies at a
distance
$\|\sigma-\sigma_{exact}\|_{L^2}=0.07\|\sigma_{exact}\|_{L^2}$ for $m=1$ and at $\|\sigma-\sigma_{exact}\|_{L^2}=0.09\|\sigma_{exact}\|_{L^2}$ for $m=2$ from
the true function. We see that in the both cases the $L^2$ norm of
the difference between the target and the model has been reduced
significantly. It is interesting to note that these results are quite similar and that it appears that little can be gained by using more patterns. In our method the interest of using extra current patterns is in the calculation of a good model function $Y_{mod}$.
\medskip

\noindent {\bf  8.The notch of the $\epsilon_0$ graph}

\noindent
We finally  discuss an  example, in which the
position and/or orientation of the target function is described by a small number of parameters.  Such a situation occurs for example in land mine detection, where the type of objects we are seeking (i.e. the model function) is quite well known and it is their location which we wish to determine.
This example takes into consideration the
action of some group of operators acting on the model function. In practice we shall encounter mainly the action
of the Euclidean group of rotation and translations. 
\begin{figwindow}[0,r,%
{\epsfig{file=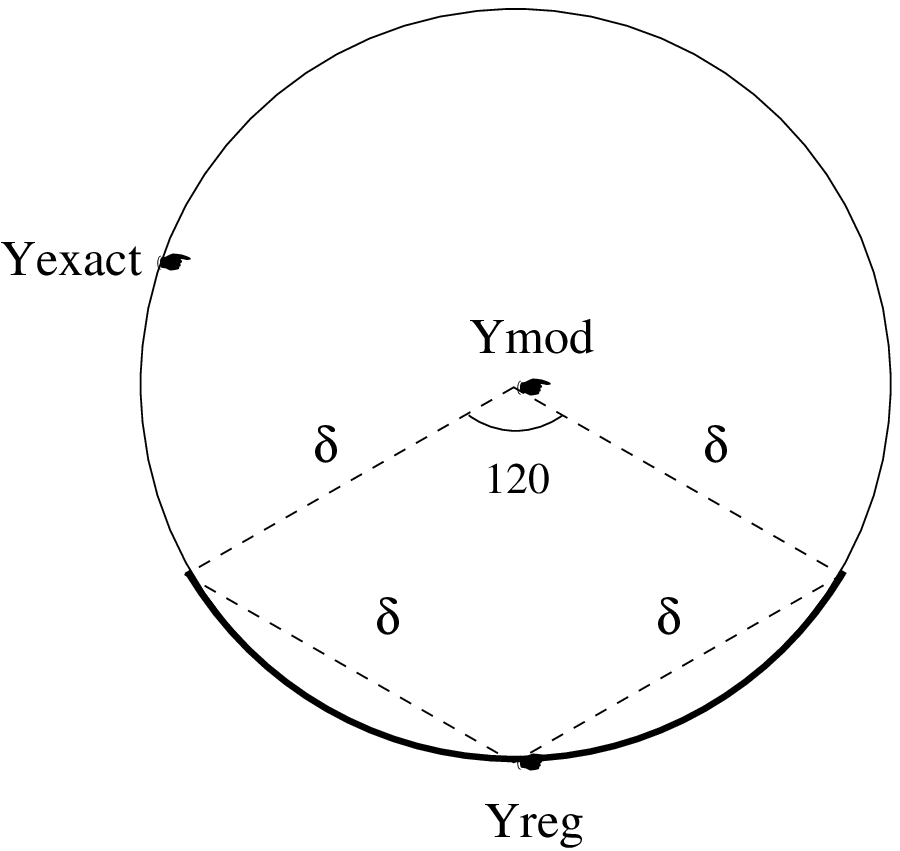,height=4cm,width=4.5cm}},%
{In two dimensions the probability that $\|Y_{reg}-Y_{exact}\|_{L^2}<\|Y_{mod}-Y_{exact}\|_{L^2}=\delta$ is only $1/3$ as  $Y_{exact}$ needs to lie on the $120^0$ bold face arc around $Y_{reg}$.} \label{probfig}]
\noindent
An interesting feature of this kind of problem is that we are no longer in the ball regularization geometry where one often encounters the situation in which the final solution does not represent an improvement with respect to the model function. This arises in simple ball regularization since the fact that both the target function  $Y_{exact}$ and the solution
$Y_{reg}$ lie on the surface of an infinite dimensional sphere of radius
$\delta$ centered at $Y_{mod}$ does not necessarly imply that $\|Y_{reg}-Y_{exact}\|_{L^2}<\|Y_{mod}-Y_{exact}\|_{L^2}=\delta$ (see Figure \ref{probfig}). In two dimensions we can see that the probability that this is true is only
 $1/3$ as  $Y_{exact}$ needs to lie on the
$120^0$ bold face arc around $Y_{reg}$,  while, in the infinite dimensional
case this probability reduces to $1/4$. This is certainly disappointing, but this situation will no longer hold if $Y_{mod}$ depends on some
unknown parameter(s) $\alpha$ so that the geometry of the regularization set is no longer spherical. We then try to find the optimal value $\alpha_0$ of the parameter $\alpha$
(see Section 5) by plotting the graph of $\epsilon_0(\alpha)$ defined in  equation (\ref{delta}). 
\end{figwindow}
\vskip -0.7cm
\begin{figure}[H]
\centering
\mbox{\subfigure[{\rmX The translation of $\sigma_{mod}$ towards $\sigma_{exact}$ for different values $\alpha$.} ]{\epsfig{figure=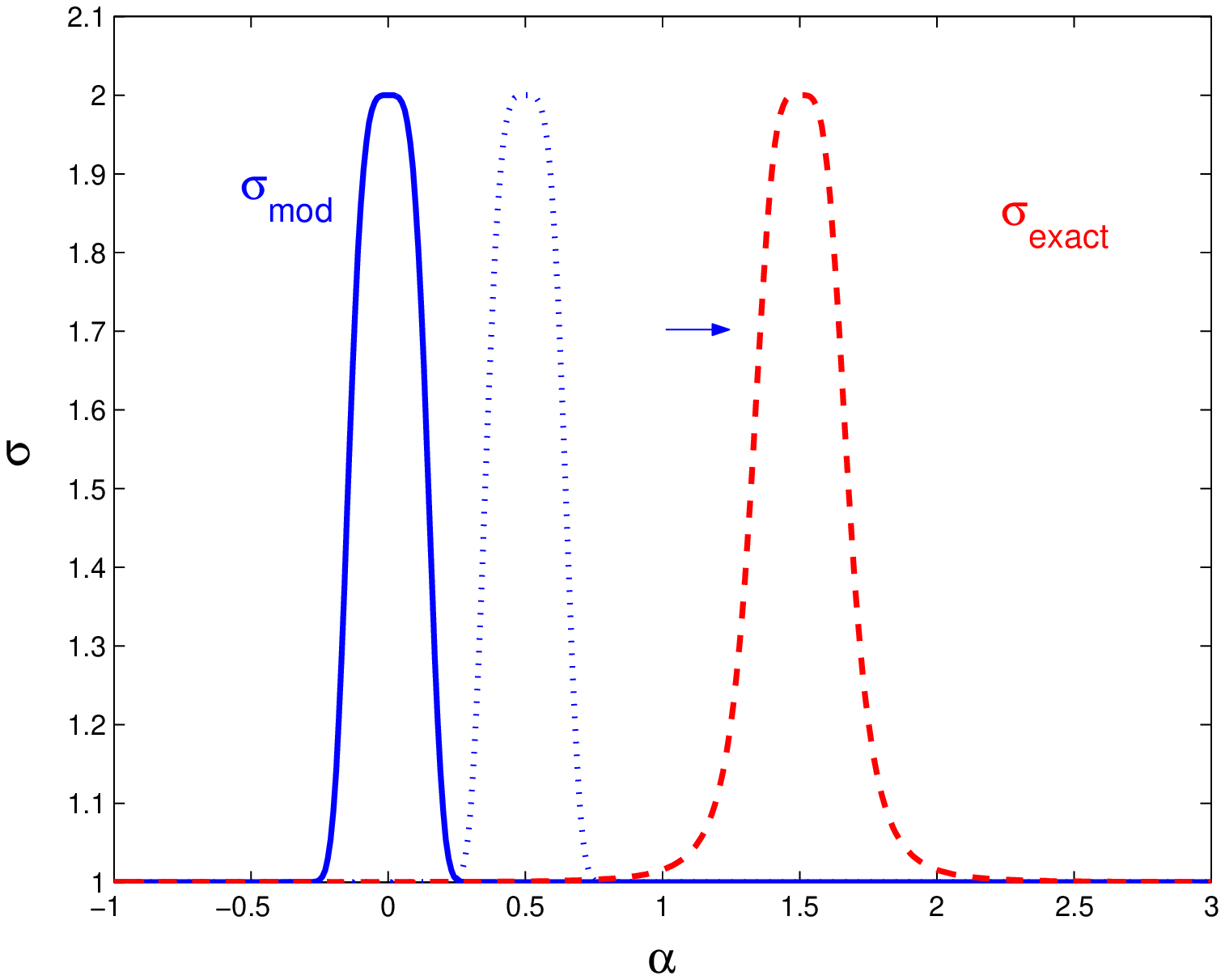,height=4.5cm, width=10cm}}\quad
\subfigure[{\rmX The notch of the graph $\epsilon_0(\alpha)$ for two different $\delta=\|Y_{reg}-Y_{exact}\|_{L^2}$} ]{\epsfig{figure=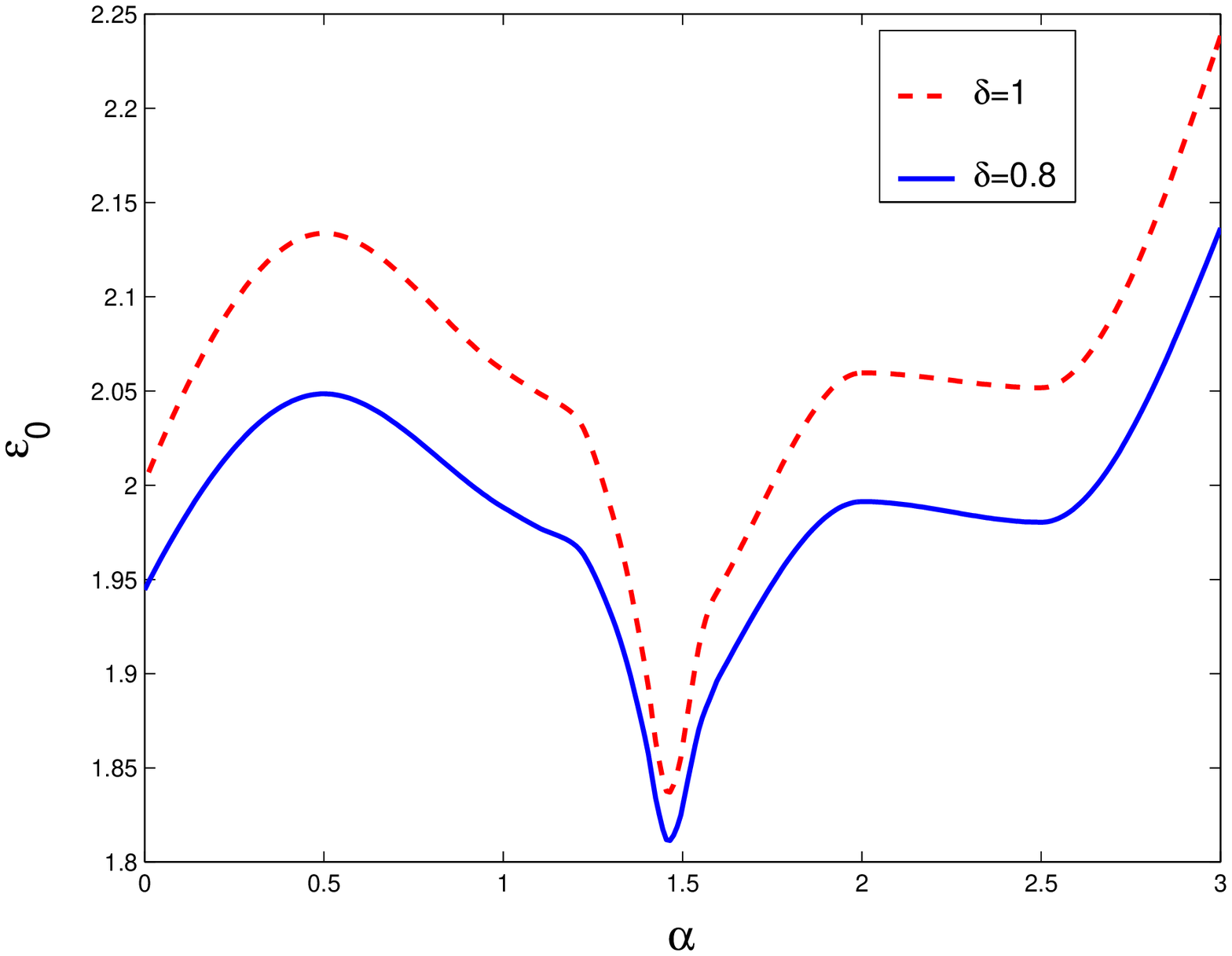,height=3.5cm, width=6cm}}}
\caption{\rmX The notch method.}
\label{notchfig}
\end{figure}
\noindent
As an example, we consider the problem of the lower half plane in which the conductivity depends on a parameter $\alpha$. The target conductivity has a maximum on the $y=-1$ axis at a point corresponding to $\alpha_0=1.5$. To illustrate how this procedure works we consider only the translations of the model function along the line $y=-1$, for $\alpha\in [0,3]$ (see Figure \ref{notchfig} (a)). We represent the target conductivity distribution by
$$\sigma_{exact}=1+\frac{a}{((x-\alpha_0)^2+(y+1)^2)^2+b}$$
and we choose the model function of the form
$$\sigma_{model}=1+\exp(-c((x-\alpha)^2+(y+1)^2)^2)\,.$$ We chose $a=b=0.001$ and $c=1500$.
\begin{figure}[ht]
\centering
\mbox{\subfigure[{\rmX The regularized conductivity $\sigma_{reg}$.}]{\epsfig{figure=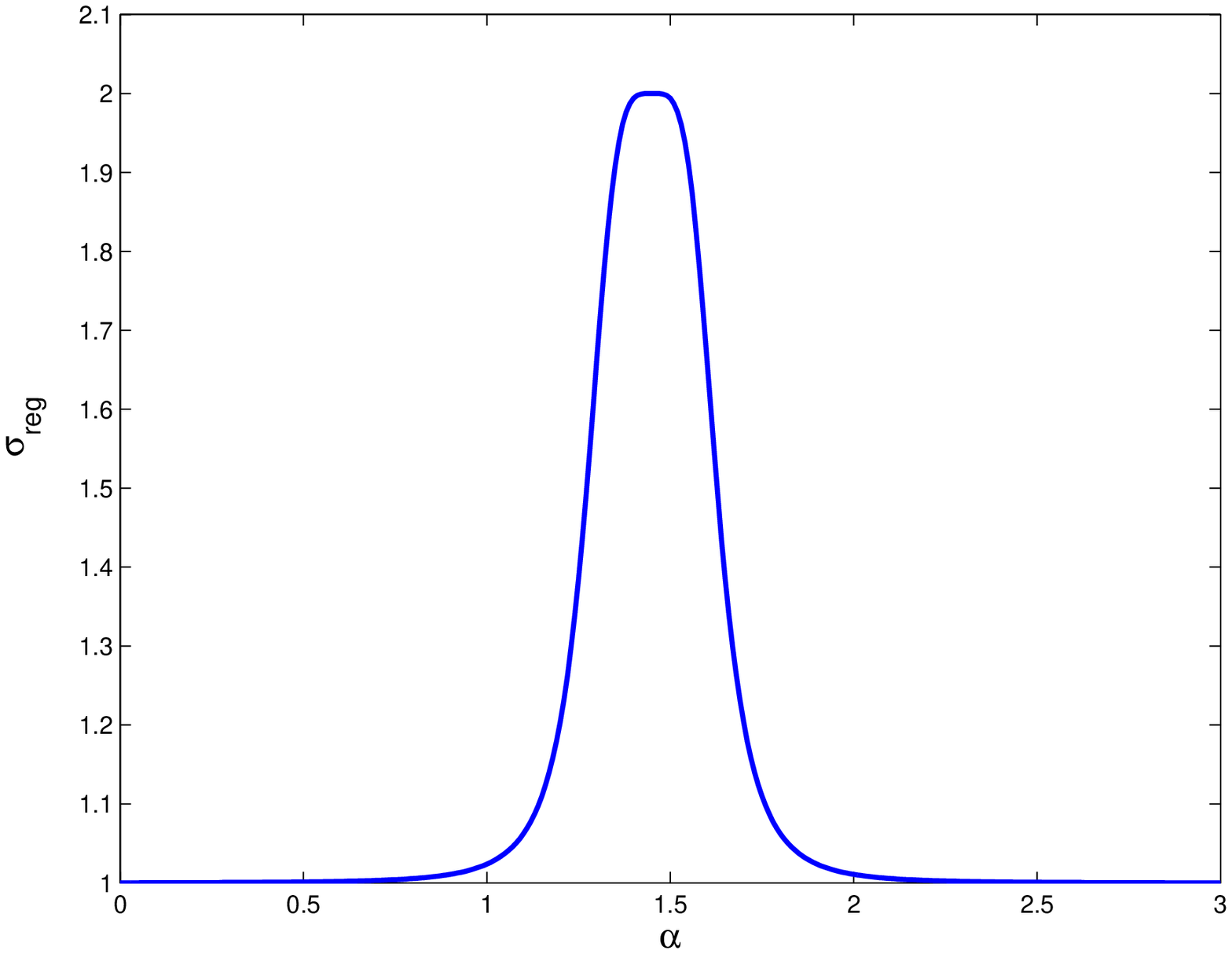,height=3.5cm, width=7cm}}\quad
\subfigure[{\rmX The absolute errors $|\sigma_{reg}-\sigma_{exact}|$.}]{\epsfig{figure=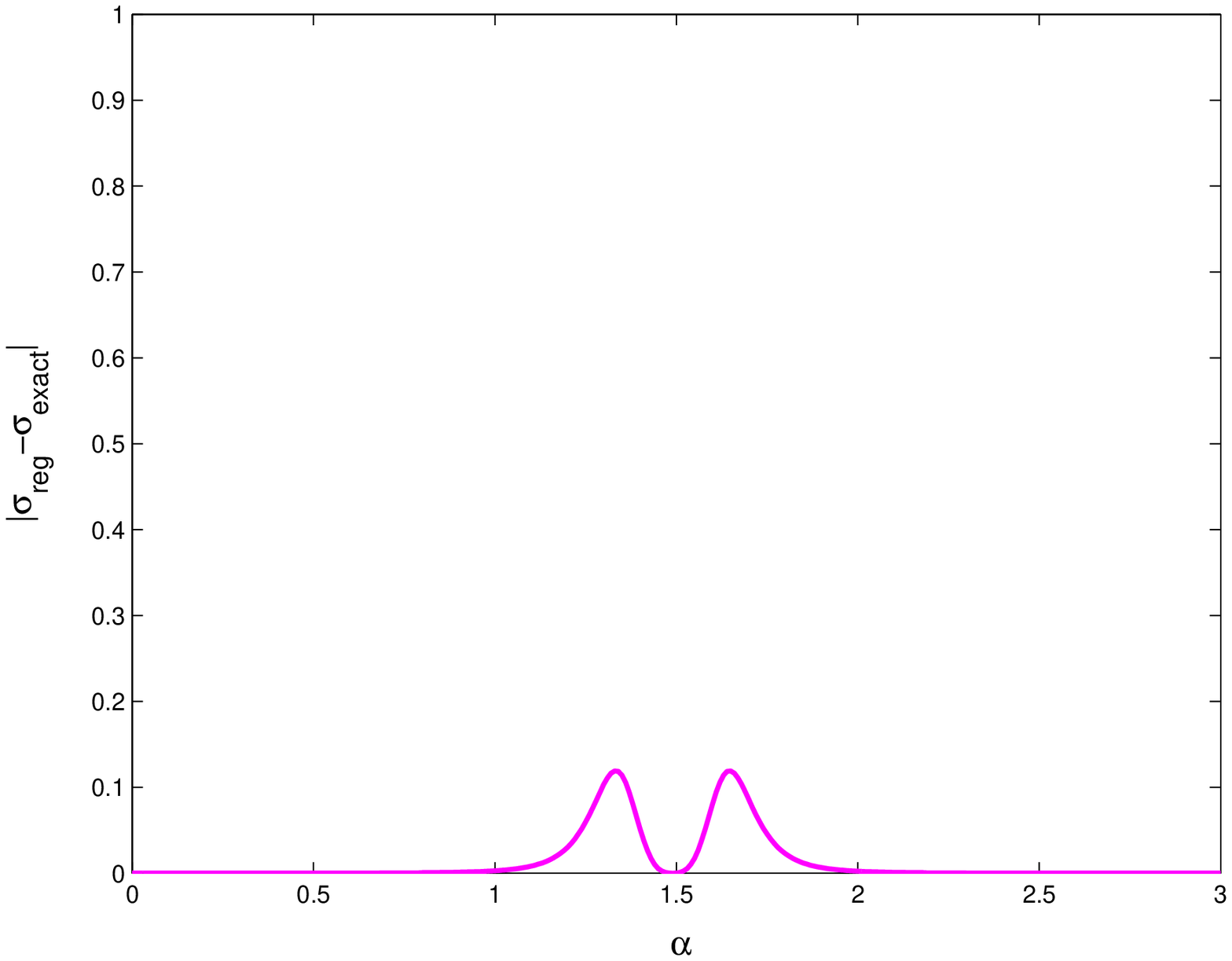,height=3.5cm, width=7cm}}}
\caption{{\rmX The reconstructed conductivity using the notch method.}}
\label{topfig}
\end{figure}

\noindent
By means of a conformal mapping we transfer the problem onto the unit disk and proceed as described earlier for different values of $\alpha$. Since this is only an illustrating example, in the regularization functional (\ref{delta}) we used the unweighted $L^2$ norm for the unit disk which means that for the initial half plane we used the norm induced  by the Jacobian of the inverse mapping. This norm has compactifying properties but it depends on the conformal mapping. Even in this rough treatement we see that this method is quite sensitive since the notch in the graph of  $\epsilon_0(\alpha)$ is quite narrow (see Figure \ref{notchfig} (b)). The position of this minimum of the graph $\epsilon_0(\alpha)$ gives the location of the object we look for and represents the best value of $\alpha$ to use in our model. The conductivity distribution reconstructed using this method is shown in Figure \ref{topfig}.

\medskip
\noindent {\bf  9. Conclusion}

\noindent
We have presented a reconstruction algorithm based on a Tikhonov regularization of a formulation of the inverse conductivity problem in terms of integral equations. The method uses prior information in the form of a model conductivity distribution. Our numerical experiments have shown the importance of this model and that multiple data appear to have little effect on our results. We have suggested ways in which more data measurements may be used to improve the model but numerical results suggest that our technique may prove to be useful in situations in which complex data collection is difficult but we have a good qualitative information on the conductivity distribution .

\medskip
\noindent
{\bf Acknowledgements}: We would like to thank Guy Auberson,
G\'erard Mennessier and Valentin Poenaru for useful conversations
concerning this work, and  The  Leverhulme Trust for providing financial
support.

\end{document}